# Analysis of Coauthorship Network in Political Science using Centrality Measures

Adeel Ahmed[1]
Department of Computer Science
National University of Modern Languages
Islamabad, Pakistan

Muhammad Fahad Khan[2]
Department of Software Engineering
Foundation University, Islamabad
Pakistan

Muhammad Usman[3], Khalid Saleem[4]
Department of Computer Science
Quaid-i-Azam University
Islamabad, Pakistan

***Abstract*—In recent era, networks of data are growing massively and forming a shape of complex structure. Data scientists try to analyze different complex networks and utilize these networks to understand the complex structure of a network in a meaningful way. There is a need to detect and identify such a complex network in order to know how these networks provide communication means while using the complex structure. Social network analysis provides methods to explore and analyze such complex networks using graph theories, network properties and community detection algorithms. In this paper, an analysis of coauthorship network of Public Relation and Public Administration subjects of Microsoft Academic Graph (MAG) is presented, using common centrality measures. The authors belong to different research and academic institutes present all over the world. Cohesive groups of authors have been identified and ranked on the basis of centrality measures, such as betweenness, degree, page rank and closeness. Experimental results show the discovery of authors who are good in specific domain, have a strong field knowledge and maintain collaboration among their peers in the field of Public Relations and Public Administration.**

## I. Introduction

Many problems in computational sciences like neuroscience, neuro-informatics, pattern recognition, signal processing and machine learning generate massive amounts of multidimensional data with multiple aspects and high dimensionality. Data is growing rapidly, day by day, because this is collected by cheap and numerous information sensing. The real world is full of different kinds of complex networks. The complexity of these networks is rapidly increasing day by day, for the enhancement and advancement in the technology. One prominent example of these type of networks is the network of internet users. According to [1], the internet users grew many fold in recent era. During last decade, from 2005 to 2015, internet users increased from 1 billion to 3.17 billion, showing the rapid growth of users. Social network analysis provides methods to explore and analyze such complex networks using graph theories, network properties and community detection algorithms. Combination of edges and nodes make a network or graph [2]. There are various types of graphs based on their characteristics. For example, the edges of facebook are undirected as shown in figure 1(b), while edges of social network of twitter are directed as shown in figure 1 (a) [3]. A graph that has some weight on its edges, is called weighted directed graph or weighted undirected graph as shown in figure 1 (c) [23,24].

The social network analysis has been widely explored to discover relationship patterns or communication patterns among individuals, teams, groups, societies, communication devices and even among organizations. The study discloses patterns of association that help in best decision making and better understanding of various patterns or groups in a graph [4].

One of the kind of social networks is coauthorship network. By applying social network analysis techniques we can discover different patterns of collaboration among authors. We can discover most active researcher, who is prominent in the field by applying different measures of social network [5]. Citation network is established, if one author cites the paper of other author and in result we obtain the network of coauthorship [6]. When author publishes a paper with another author then they form one-to-one relationship. If author has a publication with multiple co-authors then they form one-tomany relationship. And if co-authors have contributed in more than one papers then the relationship is many-to-many.

Centrality is computed by using centrality measures on directed or undirected graph. Some commonly used centrality measures are: degree centrality [5,7,8,9], closeness centrality [5,7,8,9], betweenness centrality [5,8,22] and PageRank [10,11,12,22].

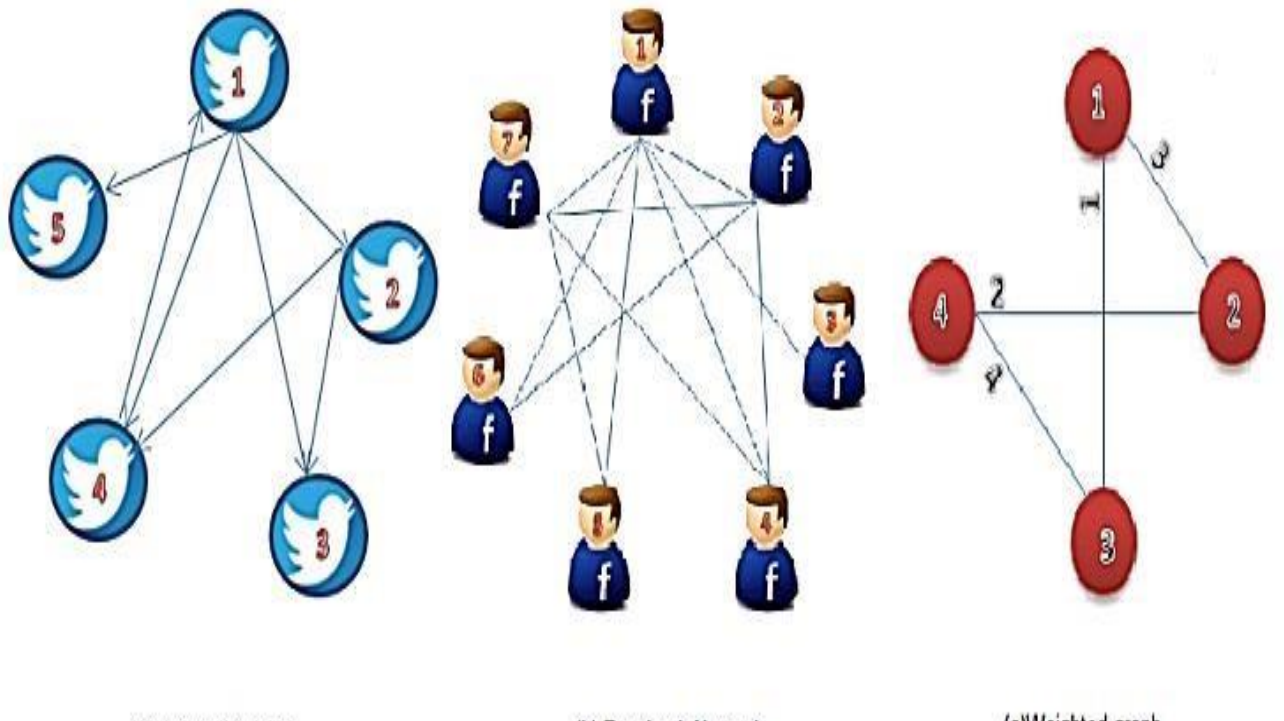

Fig. 1. Directed Twitter Network (b) Undirected facebook Network (c) Weighted Graph.

## II. Related Work

Modularity divides a complex network into small groups called modules. If the modularity value of a graph is high then it means that modules are cohesive and are strongly connected with each other. Shang et al. [13] proposed a MIGA-Modularity and Improved Genetic Algorithm to overcome the difficulty for finding optimal solution when handling large scale network problem with hill climbing. MIGA has low computational time and can detect more than half part with prior information using simulated annealing method.

Sutaria et al. proposed a community detection algorithm in which author finds the communities on the basis of modularity class [14].

Newman proposed CNM algorithm, discovering nonoverlapping and overlapping communities [15]. Palla et al. described cumulative distribution functions P(scom), P(dcom), P(sov) and P(m) that used four basic quantities. Each node represented as i of network characterize as membership number $n_i$ of the community. Communities are represented as α and β, that share overlapping property depicting the size of the community. Palla et al. used k-clique method for finding communities in a network. The benefit of this method over divisive method and agglomerative method is that it allows construction of unconstrained network of communities [16].

Karsten et al. used a simple approach with common neighboring similarity, topological clustering coefficient similarity and node attribute similarity using directed weighted graph. Proposed approach identified the clustering coefficient of the node, using clustering coefficient similarity. It measures the contribution of the connectedness among the neighboring nodes. Common neighboring similarity captures overall connectedness between immediate neighbors of nodes by substituting the neighbors. Finally, node attribute similarity computed the weight of edges based on node attribute similarity [17].

Yang et al. proposed an approach that utilized the spectral clustering algorithm which compared network communities quantitatively. Thirteen different communities were examined and divided into four classes. This methodology used for comparing networks, based on real data and examining their robustness [18]. Authors found that this method reliably detected ground-truth communities.

Qiu et al. proposed a ranking algorithm called ocdRank for finding overlapping communities in social network. The algorithm combines the features of overlapping community detection and community member ranking in heterogeneous social networks. Results show that ocdRank has low time complexity and detected better community structure as compared to other community detection methods [19].

In [20], Altunbey et al. proposed an algorithm called Parliamentary Optimization Algorithm (PAO) for finding overlapping communities in social networks.

In 2013, Li et al. embedded the six social capital measures, closeness, degree, betweenness, team exploration, profilic coauthor count and publishing tenure, for analyzing the research impact. The dataset consists of more than hundred scholars between the time span of 1999 to 2003. Li et al. analyzed the impact of social capitals on citations. Author defined the three social capital dimensions of relational, structural and cognitive capital, for coauthorship network. The results show that the „relational capital‟ and „team exploration‟ have no direct impact on citation count but „betweenness‟ has indirect effect [6].

Newman et al. performed case study on coauthorship network [5]. Author collected data from bibliographic resource, consisting of 1589 researchers as nodes and 2742 links, drawn by edges. The authors are ranked by applying four common centrality measures.

Liu et al. performed analysis on dataset using binary undirected network model [8]. The data is collected from IEEE and ACM conferences. A new network is introduced, named „weighted directional network model‟. Another dataset is obtained from ACM DL and JCDL and DBLP for IEEE ADL. This dataset contains 1567 authors, 3401 links among authors, and 759 publications. The largest component from network is observed and analysis showed that SIGMOD, NCSTRL and JCDL network have 60%, 57.2%, and 32.7% values of all authors, respectively. The results also show that DLS domains are strongly linked with scientific domain.

Yun et al. performed analysis by using micro-level properties on co-authorship network. The dataset contains information about sixteen journals from time span of 1988 to 2007. Four centrality measures that are closeness centrality, degree centrality, PageRank and betweenness centrality are used to rank top 30 authors and shows the highest collaboration among authors [21].

## III. Proposed Methodology

The proposed analysis methodology consists of three steps: First, the data is collected from Microsoft Academic Graph (MAG), then in second step, the data is preprocessed and transformed in required form, thirdly, we applied centrality measures and ranked the authors related to each field. We have chosen two fields of Political Science, Public Relations and Public Administration, and analyzed these fields using most common centrality measures. In the study, the goal is set to find most prominent group of authors in each field and ranked these authors according to work in their respective field. The proposed methodology is applied one by one on each field, which is discussed in subsequent sections.

## IV. About Dataset

Table I gives the data statistics related to the sub fields of Political Science that is Public Relations and Public Administration. Microsoft Academic Graph (MAG) is an open dataset of coauthorship network provided by Microsoft. This coauthorship network dataset is downloadable from Microsoft website. The dataset comprises of information of all aspects of the research papers including Journal, Conference and CERN and other projects. In coauthorship network, there is collaboration of co-authorship with an appropriate affiliation. Most of the publications of MAG have 2 to 15 co-authors and in some cases 6,000 co-authors, More than 30 million publications have 2 to 15 co-authors. The most productive research year for the field of Political Science, was 2013.

TABLE I. DATA STATISTICS RELATED TO PUBLIC RELATIONS AND PUBLIC ADMINISTRATION

| | Public Relations | Public Administration |
|---|---|---|
| Number of authors | 83516 | 238385 |
| Modularity | 0.999 | 0.974 |
| Network diameter | 41 | 34 |
| Connected components | 18862 | 49787 |
| Avg. clustering coefficient | 0.915 | 0.877 |
| Avg. path length | 12.834 | 24.831 |
| Avg. degree | 2.683 | 4 |

## V. RANKING AUTHORS IN PUBLIC ADMINISTRATION ON THE BASIS OF CENTRALITY MEASURES

For the analysis, common centrality measures of social networks have been applied, such as closeness centrality, degree centrality, betweenness centrality and PageRank. These metrics are used to rank authors according to their fields.

### A. Ranking Authors based on Degree Centrality

The degree centrality measure is used to find highest degree node. The degree centrality measure highlighted those scientists who have highest collaboration. The average degree distribution of public relations is 2.683. Most of the researchers have low degree and few researchers have high degree as shown in Table II.

The author named as „14674B35-DanckerDLDaamen" of public relation affiliated to Leiden University, has highest influence and frequent collaboration with other 47 researchers as shown in figure 2. „14674B35-DanckerDLDaamen" has worked exclusively in public opinion field which is the sub field of public relations. The second most influence author is „7FF2291D-DarrelMontero" and is affiliated with Arizona State University.

We extracted the graph of top 10 degree researchers and their connected researchers as shown in figure 3. This graph contains 426 researchers and 1146 collaborations. Average degree of top 10 degree graph is 5.38, network diameter is 4, modularity is 0.7 and there are 11 connected components in the network. Modularity value shows that this graph has good community structure. In figure 5, the most productive institute is the Univeristy of Missouri. „7F4328BD-GlenTCameron" is the researcher who has degree 38 and ranked as 4$^{th}$ in top ten degree, with 41 other researchers. The author collaborated with University of Missouri, Missouri School of Journalism and University of Georgia and he has productive research with University of Missouri as he has 19, 6 and 1 publications, respectively. The second most productive institute is University of Minnesota. „7E654E5D-DavidPFan" is the researcher who has 27 degree and ranked as 10 in top ten degree researchers, having collaboration with 28 other researchers. The author is affiliated to University of Minnesota and he has eleven publications.

### B. Ranking Authors based on betweenness Centrality

Betweenness centrality ranks the nodes with highest value that are part of most of the shortest path. The network diameter of public relations is 41 and the length of average path is 12.833. Majority of the researchers have zero or near to zero betweenness, some researchers have high betweenness, which shows that they are responsible for flow of knowledge from one community to another community.

TABLE II. AUTHORS RANKING OF PUBLIC RELATIONS ON BASIS OF DEGREE CENTRALITY WITH RESPECT TO OTHERS

| Author | Degree | Rank | Betweenness | Rank | Closeness | Rank | PageRank | Rank |
|---|---|---|---|---|---|---|---|---|
| 14674B35-DanckerDLDaamen | 46 | 1 | 8.85E-07 | 381 | 6.18E-04 | 1424 | 5.62E-05 | 101 |
| 7FF2291D-DarrelMontero | 44 | 2 | 5.46E-07 | 425 | 7.17E-04 | 1409 | 8.86E-05 | 19 |
| 0B211A8C-PaulSlovic | 42 | 3 | 1.28E-04 | 35 | 3.70E-03 | 108 | 1.18E-04 | 6 |
| 7F4328BD-GlenTCameron | 38 | 4 | 2.21E-04 | 12 | 3.80E-03 | 73 | 1.51E-04 | 2 |
| 2A8E03FD-SFMccool | 36 | 5 | 2.99E-06 | 308 | 7.23E-04 | 1408 | 8.31E-05 | 27 |
| 0106C2B9-RobertJBlendon | 35 | 6 | 1.13E-04 | 42 | 2.98E-03 | 650 | 8.80E-05 | 21 |
| 7D5AAC1C-FranciscoHGFerreira | 31 | 7 | 3.77E-07 | 462 | 4.55E-04 | 1548 | 4.10E-05 | 259 |
| 290A255A-JillRoessner | 29 | 8 | 1.90E-07 | 533 | 5.09E-04 | 1485 | 4.41E-05 | 208 |
| 771B6FCA-DietramAScheufele | 28 | 9 | 7.29E-04 | 1 | 4.43E-03 | 1 | 8.26E-05 | 29 |
| 7E654E5D-DavidPFan | 27 | 10 | 1.75E-04 | 23 | 3.70E-03 | 109 | 1.07E-04 | 10 |

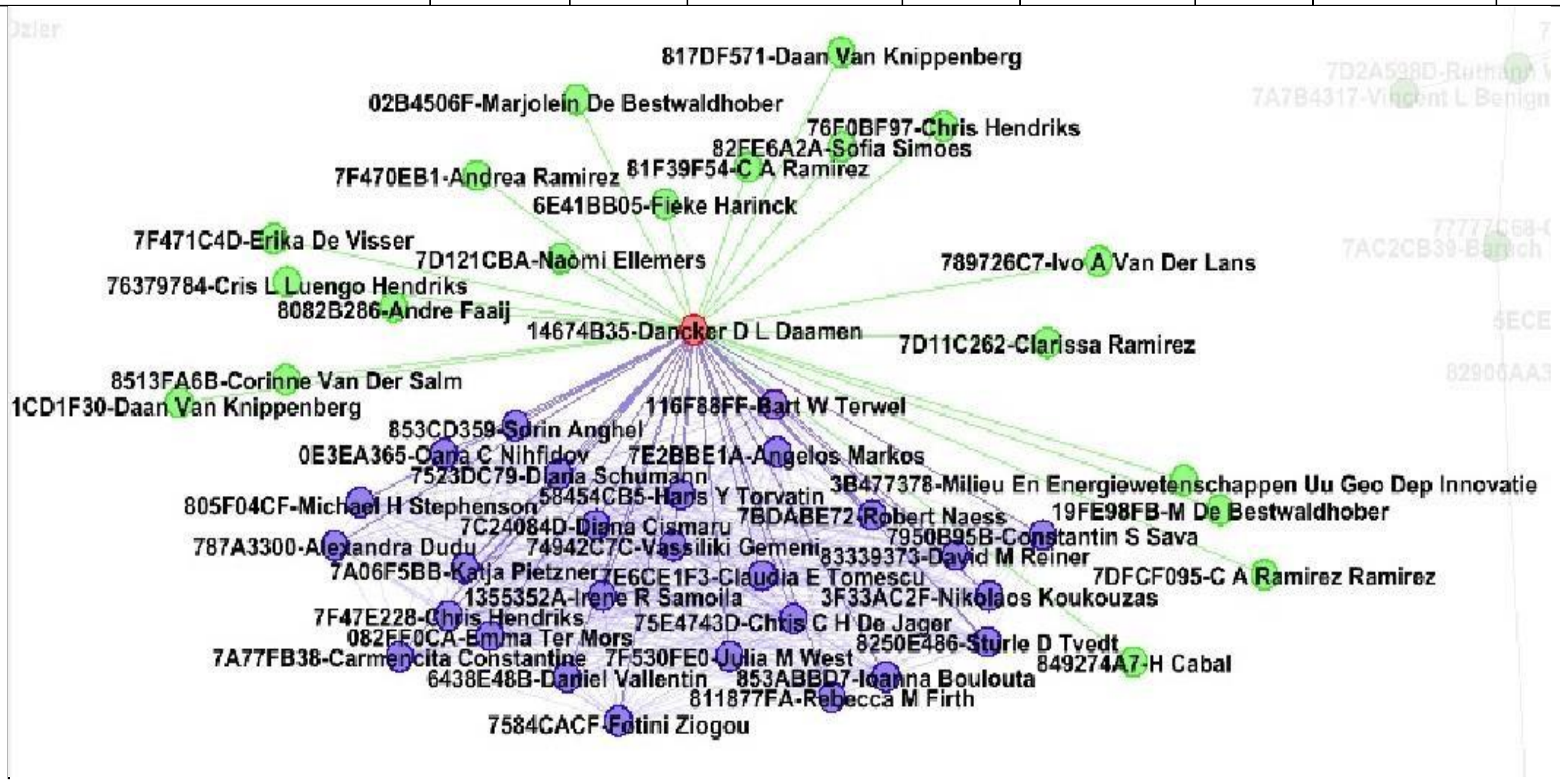

Fig. 2. „7FF2291D-DarrelMontero" with Highest Degree Centrality.

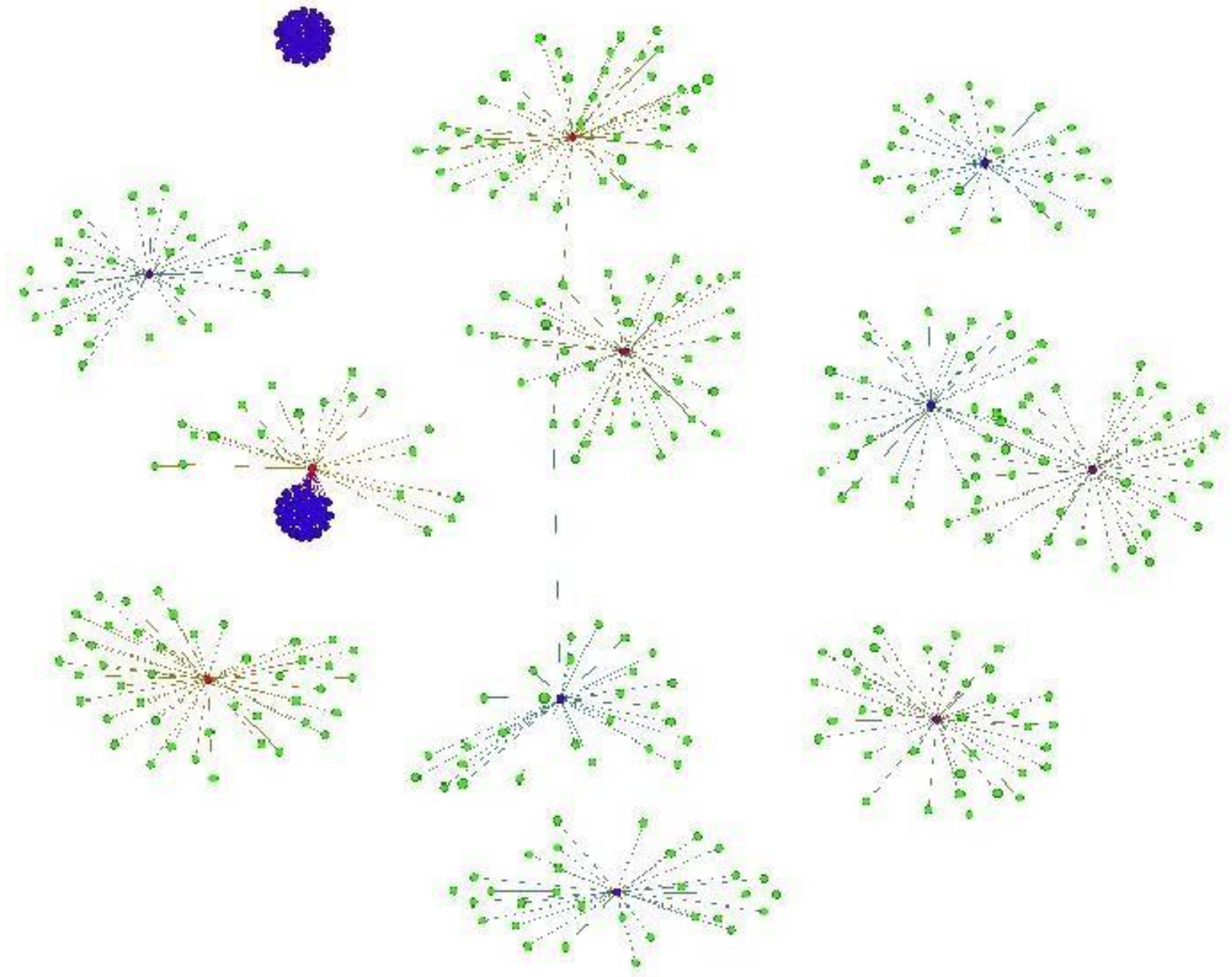

Fig. 3. Top 10 Authors of Public Relations having Highest Degree Centrality.

TABLE III. RANKING AUTHORS ON THE BASIS OF BETWEENNESS CENTRALITY

| Author | Degree | Rank | Betweenness | Rank | Closeness | Rank | PageRank | Rank |
|---|---|---|---|---|---|---|---|---|
| 771B6FCA-DietramAScheufele | 28 | 9 | 1412952.09 | 1 | 4.43E-03 | 1 | 8.26E-05 | 29 |
| 80EE5C66-JeongnamKim | 12 | 25 | 1146458.61 | 2 | 4.41E-03 | 3 | 5.26E-05 | 113 |
| 7CF2B524-DoohunChoi | 5 | 32 | 1139426.84 | 3 | 4.42E-03 | 2 | 1.98E-05 | 1851 |
| 7E3071EE-BeylingSha | 12 | 25 | 762604.15 | 4 | 4.32E-03 | 4 | 4.86E-05 | 155 |
| 7CF3C0D4-ElizabethLToth | 20 | 17 | 709156.65 | 5 | 4.22E-03 | 6 | 8.66E-05 | 23 |
| 76015751-BryanHReber | 14 | 23 | 606510.35 | 6 | 4.01E-03 | 28 | 5.01E-05 | 133 |
| 805E4884-PatriciaMoy | 12 | 25 | 530612.87 | 7 | 4.18E-03 | 7 | 5.20E-05 | 118 |
| 4AF7AF7E-KrishnamurthySriramesh | 22 | 15 | 479380.97 | 8 | 4.18E-03 | 8 | 9.75E-05 | 13 |
| 72B6EC1A-DebashishMunshi | 5 | 32 | 437850 | 9 | 3.09E-03 | 560 | 2.14E-05 | 1625 |
| 5F07A3FF-VericaRupar | 6 | 31 | 432422 | 10 | 2.89E-03 | 744 | 2.91E-05 | 743 |

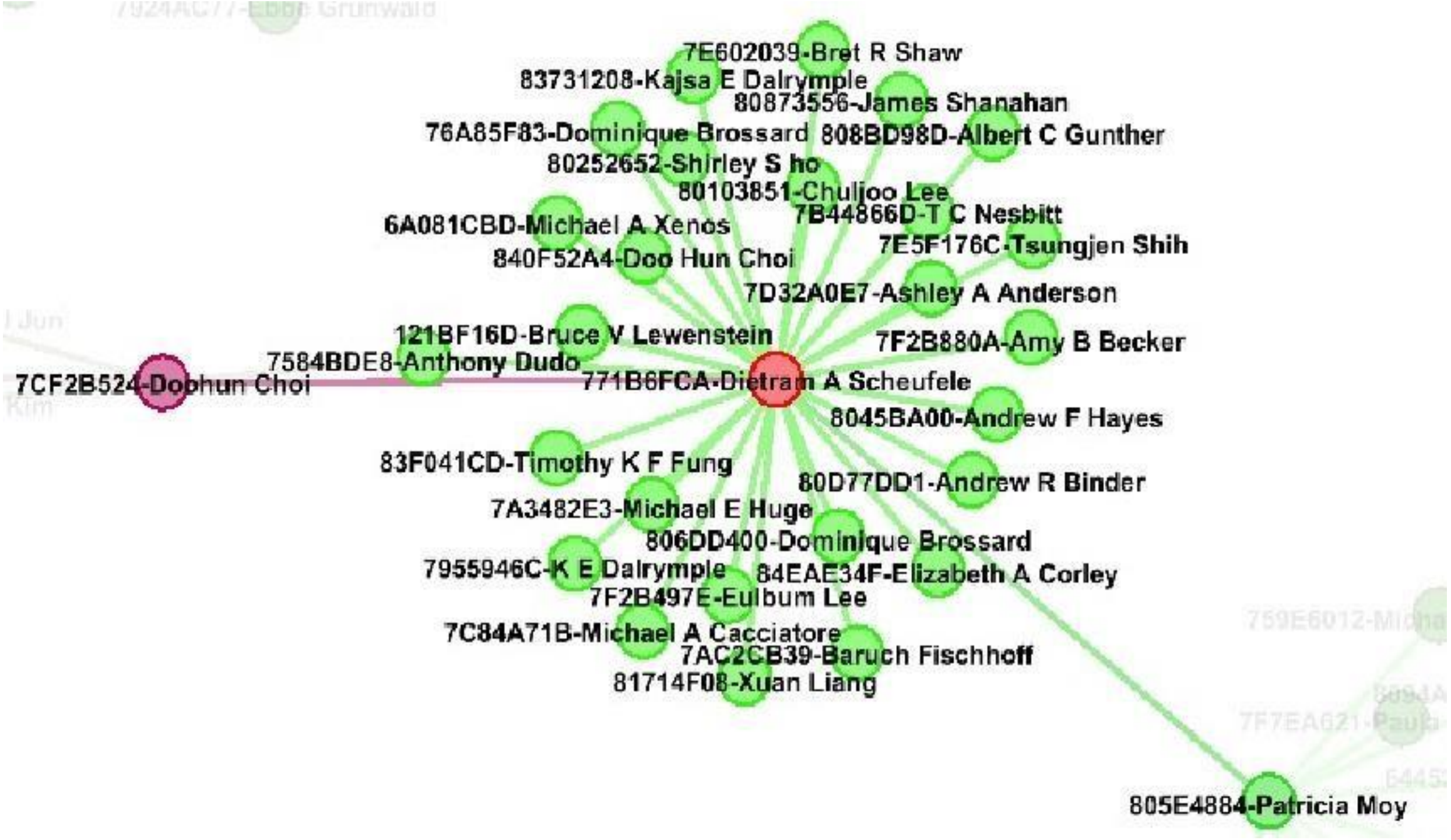

Fig. 4. „771B6FCA-DietramAScheufele‟, Author of Public Relations having Highest Betweenness Centrality.

TABLE IV. RANKING AUTHORS ON THE BASIS OF CLOSENESS CENTRALITY

| **Author** | **Degree** | **Rank** | **Betweenness** | **Rank** | **Closeness** | **Rank** | **PageRank** | **Rank** |
|---|---|---|---|---|---|---|---|---|
| 771B6FCA-DietramAScheufele | 28 | 9 | 7.29E-04 | 1 | 4.43E-03 | 1 | 8.26E-05 | 29 |
| 7CF2B524-DoohunChoi | 5 | 32 | 5.88E-04 | 3 | 4.42E-03 | 2 | 1.98E-05 | 1851 |
| 80EE5C66-JeongnamKim | 12 | 25 | 5.92E-04 | 2 | 4.41E-03 | 3 | 5.26E-05 | 113 |
| 7E3071EE-BeylingSha | 12 | 25 | 3.93E-04 | 4 | 4.32E-03 | 4 | 4.86E-05 | 155 |
| 0916F08B-JamesEGrunig | 15 | 22 | 2.06E-04 | 16 | 4.30E-03 | 5 | 6.28E-05 | 70 |
| 7CF3C0D4-ElizabethLToth | 20 | 17 | 3.66E-04 | 5 | 4.22E-03 | 6 | 8.66E-05 | 23 |
| 805E4884-PatriciaMoy | 12 | 25 | 2.74E-04 | 7 | 4.18E-03 | 7 | 5.20E-05 | 118 |
| 4AF7AF7E-KrishnamurthySriramesh | 22 | 15 | 2.47E-04 | 8 | 4.18E-03 | 8 | 9.75E-05 | 13 |
| 7584BDE8-AnthonyDudo | 2 | 35 | 0.00E+00 | 1064 | 4.18E-03 | 9 | 8.19E-06 | 5052 |
| 7F6A3D86-SeihillKim | 4 | 33 | 1.18E-06 | 364 | 4.17E-03 | 10 | 1.99E-05 | 1832 |

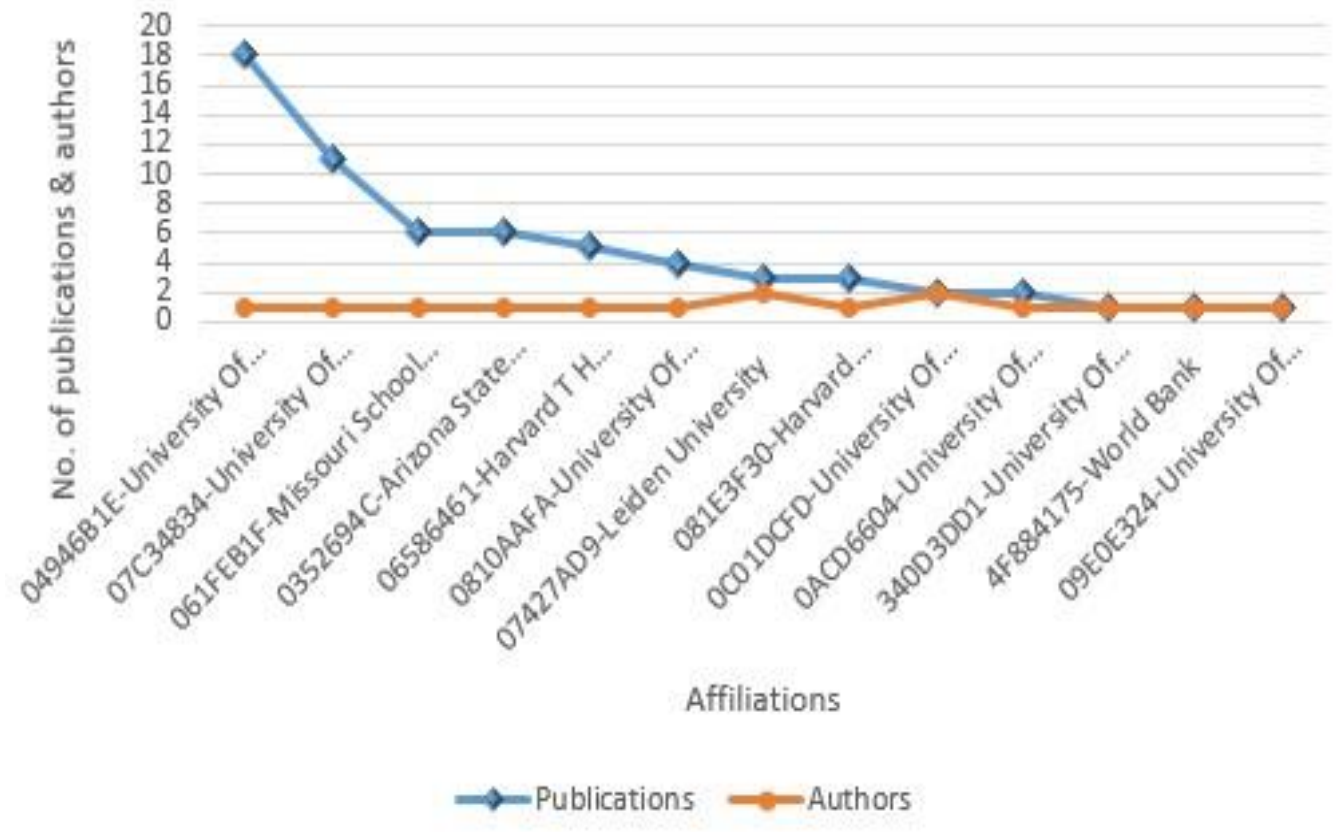

Fig. 5. Institutes and their Publications of Top 10 Researchers W.R.T Degree.

Table III shows the top 10 researchers who have high betweenness in the field of public relations. Figure 4 shows the graph that contains 119 researchers and 129 collaborations. The network diameter of graph is 8, average path length is 4.32, 0.754 modularity and there are 2 connected components. The most influential author is „771B6FCA-DietramAScheufele" who is affiliated to „University Of Wisconsin Madison", „Nanyang Technological University", „Ohio State University", „Cornell University", „University of Washington" and „University of Wisconsin Madison School of Journalism Mass Communication". The author is the most central researcher and is involved in shortest path from one researcher to other researcher and have frequent collaboration, as he is ranked 9 in degree centrality measures.

Node „80EE5C66-JeongnamKim" is the second most central researcher having frequent collaborations. He has ranked 25th in degree centrality measures, affiliated to „Purdue University", „University Of Houston", „University Of Maryland College Park", „Hankuk University of Foreign Studies",„University Of Siena", „Hong Kong Baptist University", „Indiana University", „Kansas State University" and „San Diego State University". He has worked in multiple fields like „Reputation", „Soft Power" and „News Media", subfields of public relations. He has collaborated with 13 other researchers.

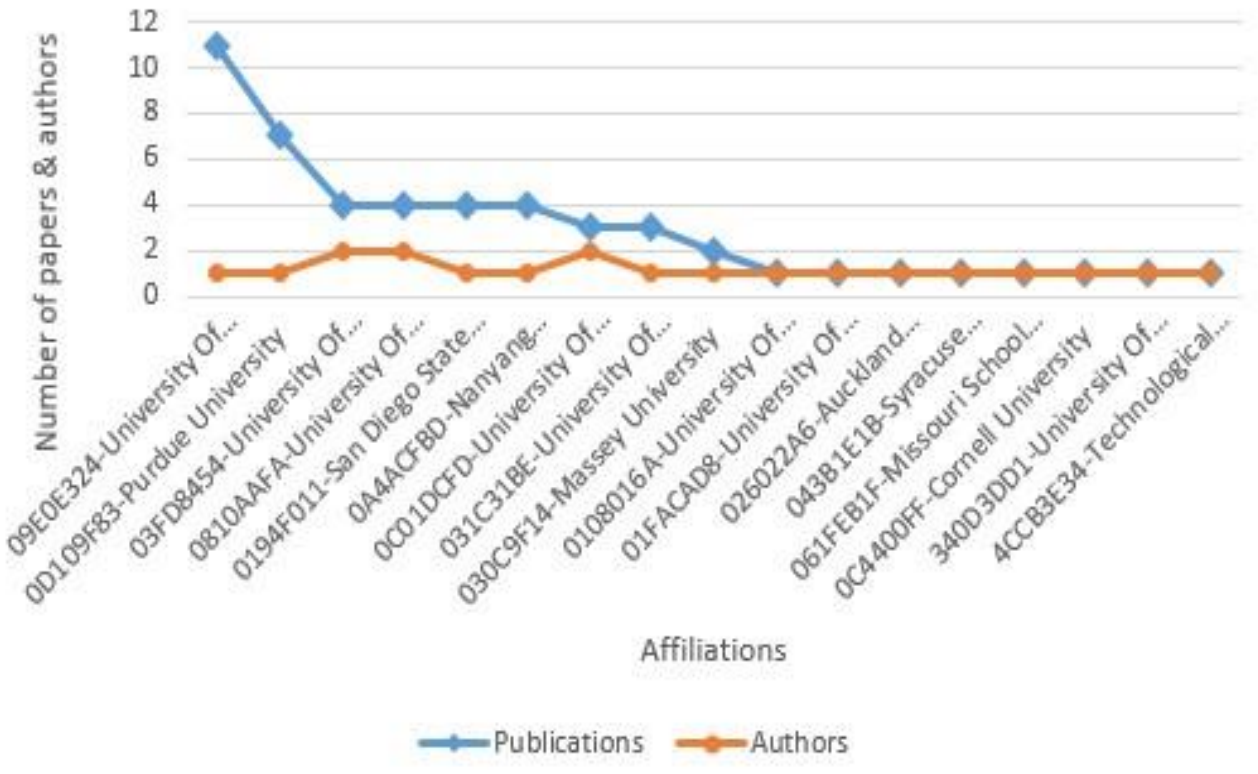

Fig. 6. Institutes and their publications of top 10 researchers w.r.t betweenness

In figure 6, the most productive institute is the Univeristy of Georgia as this institute has highest number of publications. „76015751-BryanHReber" is the researcher who has degree14 and ranked as 6th in top 10 betweenness researchers, having collaboration with 15 other researchers. He has collaboration with University of Georgia, University of Alabama, Missouri School of Journalism, University of Florida and University of Maryland College Park, as he has 11, 8, 4, 2 and 1 publications, respectively. The second most productive institute is Purdue University. „80EE5C66-JeongnamKim" is the researcher who has degree 12 and ranked at second place in top 10 betweennes researchers. having has collaboration with 13 other reasearchers, in collaboration with Purdue University, University of Maryland College Park ,University of Houston, Hankuk University of Foreign Studies, University of Siena, Hong Kong Baptist University, Kansas State University, San Diego State University and Indiana University. He has 7, 3, 2, 1, 1, 1, 1, 1 and 1 publications with these institutions, respectively.

*C. Ranking Authors based on Closeness Centrality* The author „771B6FCA-DietramAScheufele" is most central researcher and ranked first in betweenness and closeness centrality, as shown in Table IV. He has worked exclusively in public opinion field which is the sub field of public relations. „7CF2B524-DoohunChoi" is the second most central researcher. Graph of top 10 researchers based on closeness centrality is shown in figure 8. This graph contains 105 researchers and 121 collaborations. The diameter of network is 7, average path length is 4.073, 0.683 is modularity and there is a single component.

Figure 7 shows the most productive institute that is „03FD8454- University Of Maryland College Park". „0916F08B-James E Grunig", „7CF3C0D4-Elizabeth L Toth" and „7E3071EE-Beyling Sha" researchers are affiliated to „„„03FD8454-University Of Maryland College Park""" and they have 4, 3 and 1 publications, respectively. The second most productive institute is „0D109F83-Purdue University". „80EE5C66-Jeongnam Kim" researcher is affiliated to „0D109F83-Purdue University" and has 7 publications.

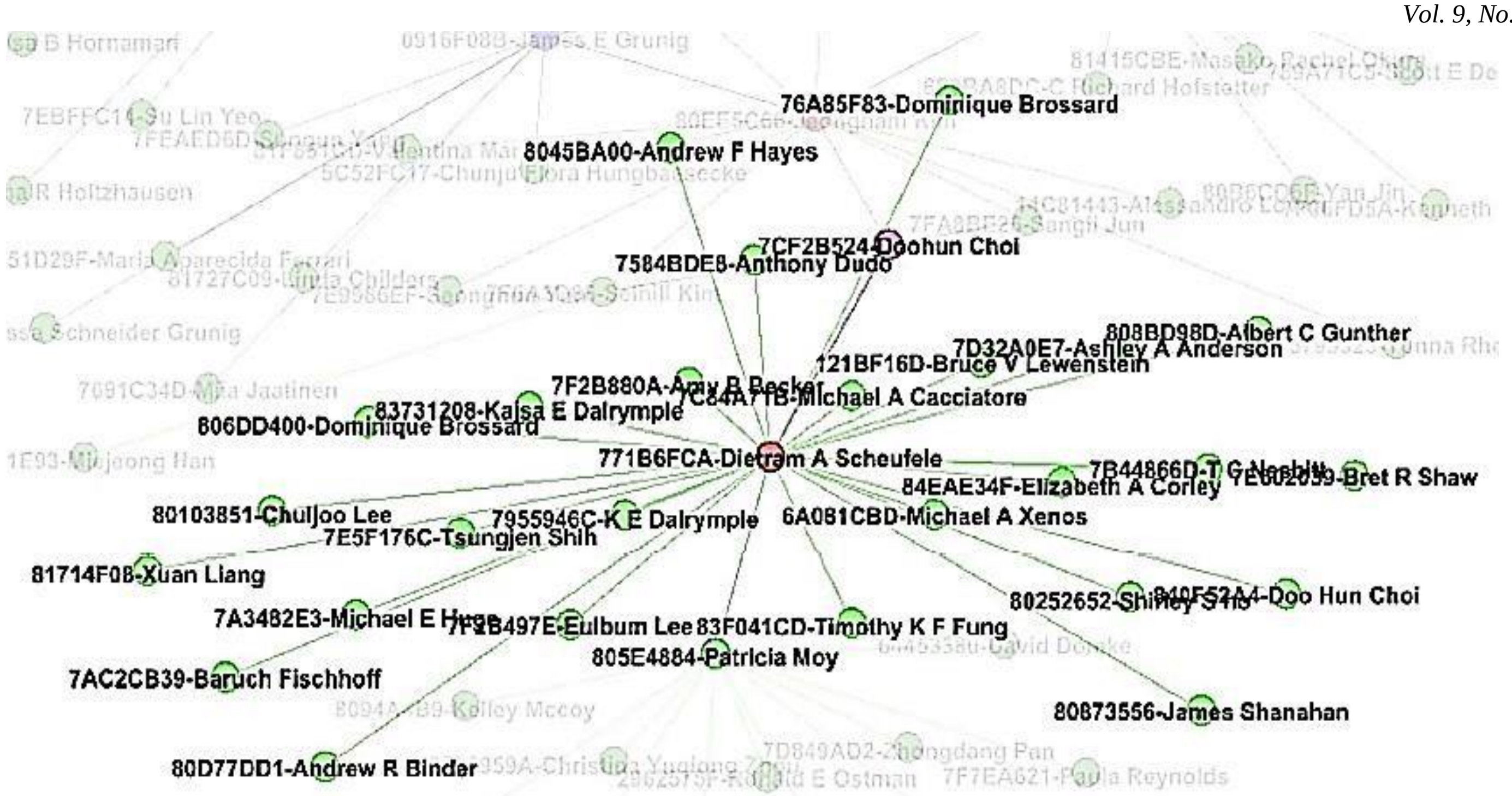

Fig. 8. Top 10 Researchers of Public Relations based on Closeness Centrality.

TABLE V. RANKING AUTHORS ON THE BASIS OF PAGERANK

| Authors/Researchers | PageRank Value | Rank |
|---|---|---|
| 7F4328BD-GlenTCameron | 1.51E-04 | 1 |
| 4AA5A185-JamesNDruckman | 1.32E-04 | 2 |
| 7CF120D6-RichardDWaters | 1.31E-04 | 3 |
| 7DE76C34-LeeBBecker | 1.18E-04 | 4 |
| 0B211A8C-PaulSlovic | 1.18E-04 | 5 |
| 81353F03-MaureenTaylor | 1.11E-04 | 6 |
| 2B74CFC5-StantonAGlantz | 1.10E-04 | 7 |
| 81A7F237-RobertLHeath | 1.09E-04 | 8 |
| 7E654E5D-DavidPFan | 1.07E-04 | 9 |
| 811A205F-WilliamLBenoit | 9.94E-05 | 10 |

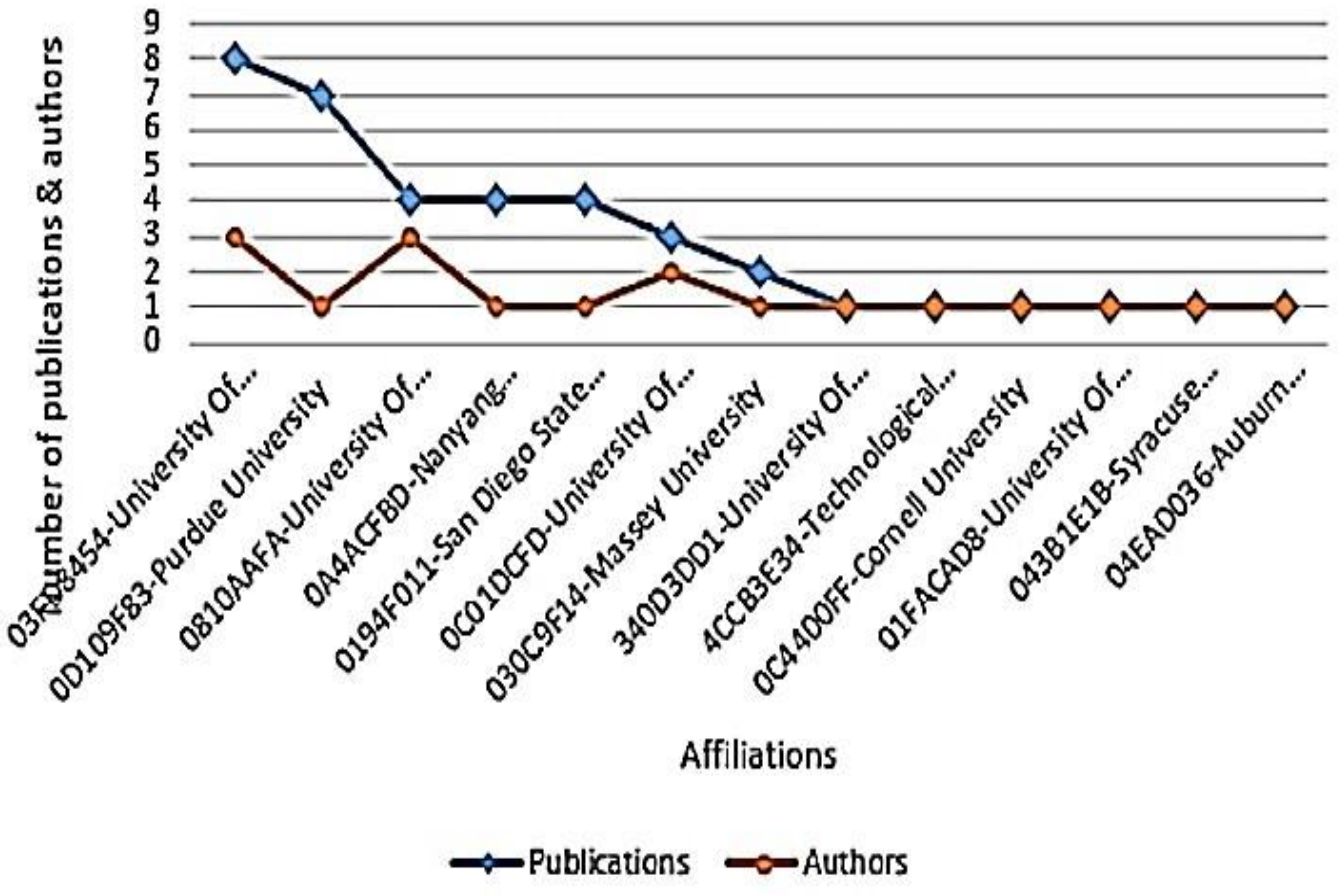

Fig. 7. Institutes and their Publications of Top 10 Researchers W.R.T Closeness.

## D. Ranking Authors based on PageRank

We have discussed top ten researchers having highest PageRank centrality of „Public Relations-025B78CE", as shown in Table V.

Figure 9 shows the researcher „7F4328BDGlenTCameron", who has the highest PageRank, and has worked in „03FEE94E-Media Relations", „09820AAE-Communication Management", „09BDF000Corporate Communication" and „071FA02B-Journalism" fields. „7F4328BD-GlenTCameron" is affiliated with three different affiliations i.e. „04946B1EUniversity of Missouri", „061FEB1F-Missouri School of Journalism" and „09E0E324University of Georgia". Figure 9 contains 281 nodes and 272 edges. Network diameter is 4, modularity is 0.889, average path length is 2.271 and there are 9 connected components.

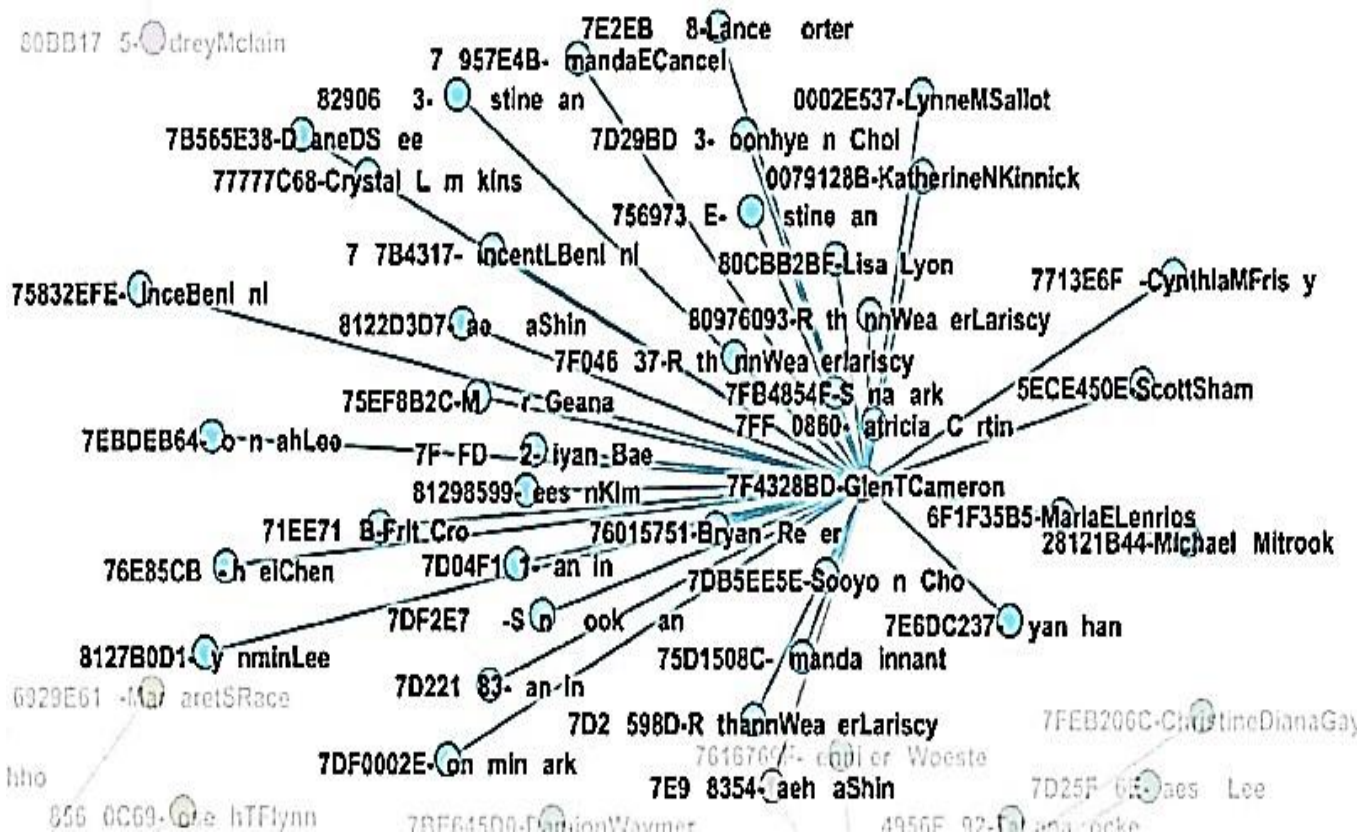

Fig. 9. Author „7F4328BD-GlenTCameron" having Highest PageRank Centrality.

The author of public administration, named as „12F4FDCC-Eds" is affiliated to „Centro Agronomico Tropical De Investigacion Y Ensenanza", who has highest influence and frequent collaboration with 110 researchers as shown in Table VI and in figure 10. „12F4FDCC-Eds" has prominent worked in 0B2F54F0-Kenya, 0A51FEF5-Refugee, 034E1111-International Law, which are the sub-fields of public administration.

The second most influencing and frequent collaborative author is „7EBE0990-RobertEBlack", affiliated to „0A183231Johns School of Public Health". He also has worked with other different affiliations i.e „05B090CE-University of California Berkeley", „08A948CC-Johns Hopkins University", „4FBCBEC0-United Nations High Commissioner For Refugees". „7EBE0990-RobertEBlack" has worked in „0A51FEF5-Refugee",„0AAE1030-Containment", „0B2F54F0-Kenya" and „063ABE50-Displaced Person" fields which are sub-fields of public administration and he has collaborated with 70 other researchers.

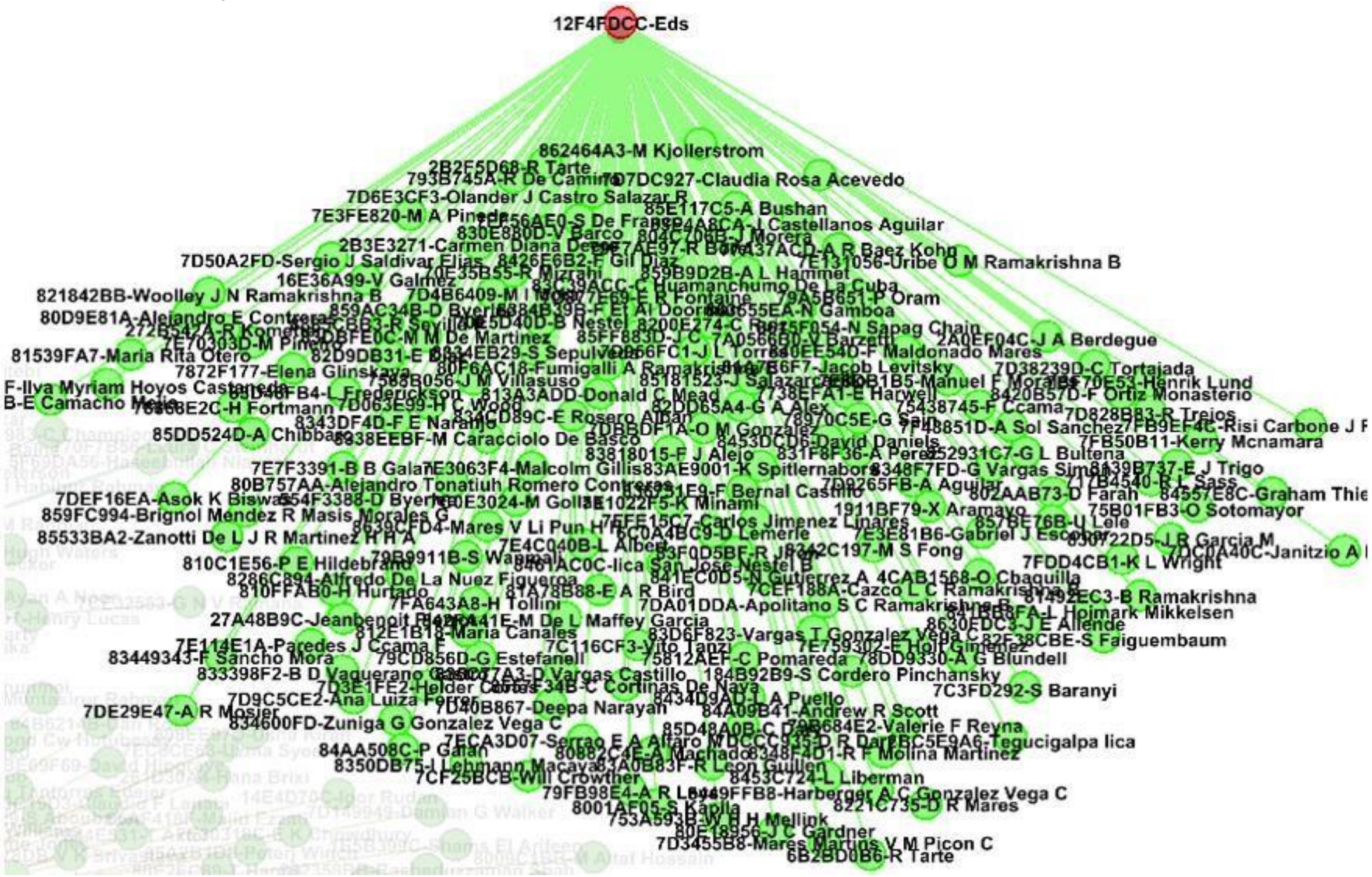

Fig. 10. „12F4FDCC-Eds", Author of Public Admininstration having Highest Degree Centrality.

## VI. Ranking Authors in Public Administration on the Basis of Centrality Measures

### A. Ranking Authors based on Degree Centrality

The average degree distribution of public administration field is 3.924. In public administration field, most of the researchers have low degree and some have high degree.

TABLE VI. RANKING AUTHORS ON THE BASIS OF DEGREE CENTRALITY

| Author | Degree | Rank | Betweenness | Rank | Closeness | Rank | PageRank | Rank |
|---|---|---|---|---|---|---|---|---|
| 12F4FDCC-Eds | 164 | 1 | 1.39E-03 | 18 | 3.73E-02 | 2535 | 5.36E-05 | 13 |
| 7EBE0990-RobertEBlack | 159 | 2 | 4.44E-03 | 1 | 4.62E-02 | 1 | 5.44E-05 | 11 |
| 0CAEADF8-Vu | 146 | 3 | 3.71E-03 | 2 | 3.72E-02 | 2674 | 8.72E-05 | 3 |
| 7C467844-FrancoisDabis | 130 | 4 | 1.07E-03 | 45 | 4.29E-02 | 24 | 3.33E-05 | 84 |
| 766E0394-ADHarries | 109 | 5 | 6.03E-04 | 152 | 4.16E-02 | 104 | 4.04E-05 | 36 |
| 8068F04B-DavidMckenzie | 103 | 7 | 2.27E-03 | 6 | 4.29E-02 | 23 | 6.43E-05 | 5 |
| 7B95835A-DavidHPeters | 99 | 8 | 3.43E-03 | 3 | 4.48E-02 | 2 | 5.50E-05 | 10 |
| 781D4EE0-ZulfiqarABhutta | 93 | 9 | 2.27E-03 | 5 | 4.40E-02 | 4 | 3.15E-05 | 108 |
| 14ABE527-DavidRBangsberg | 92 | 10 | 2.00E-03 | 8 | 4.33E-02 | 13 | 2.62E-05 | 240 |

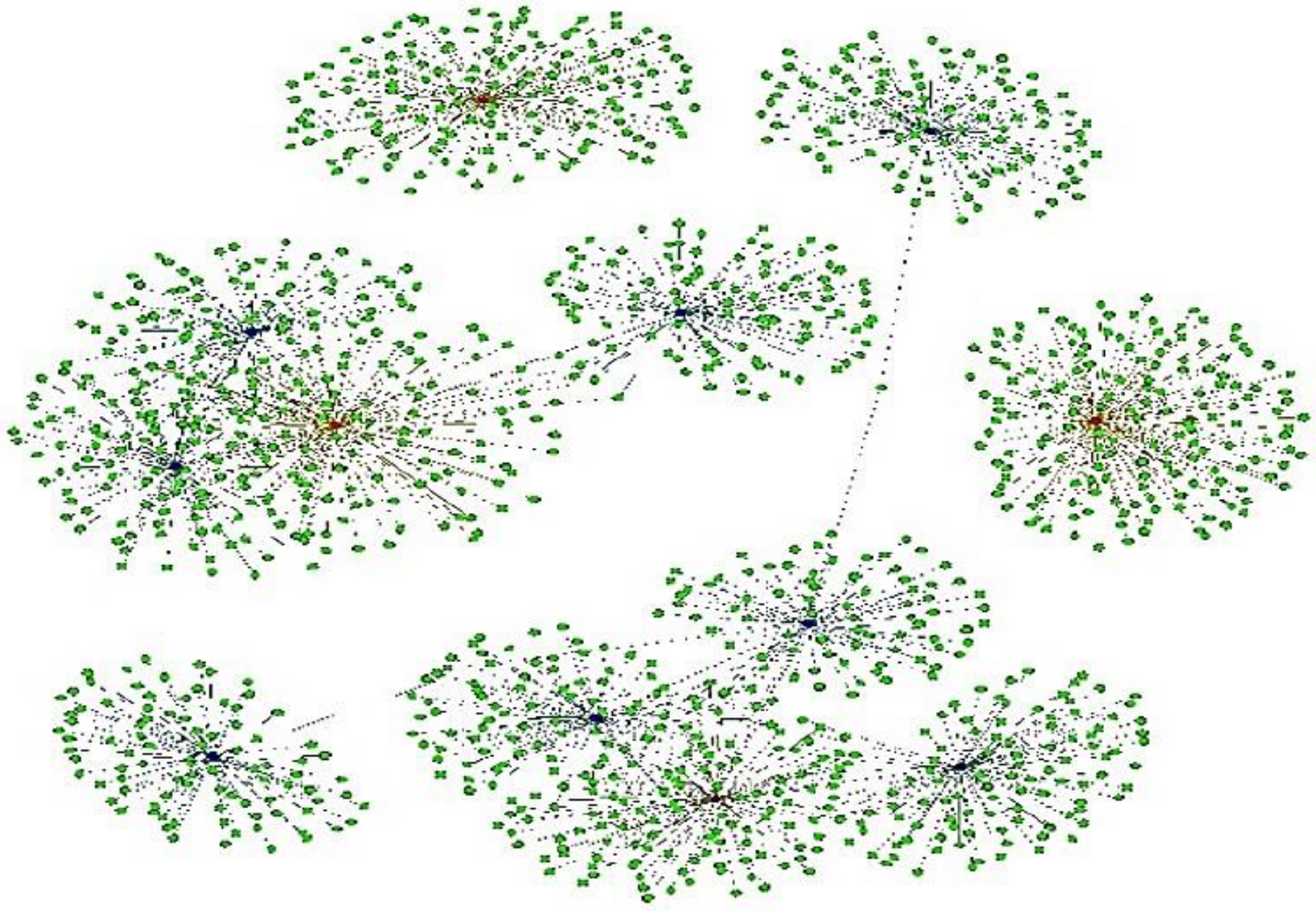

Fig. 11. Top 10 Researchers on the basis of Degree Centrality.

TABLE VII. Ranking Authors on the Basis of Betweeness Centrality

| **Author** | **Degree** | **Rank** | **Betweenness** | **Rank** | **Closeness** | **Rank** | **PageRank** | **Rank** |
|---|---|---|---|---|---|---|---|---|
| 7EBE0990-RobertEBlack | 159 | 2 | 4.44E-03 | 1 | 4.62E-02 | 1 | 5.44E-05 | 11 |
| 0CAEADF8-Vu | 146 | 3 | 3.71E-03 | 2 | 3.72E-02 | 2674 | 8.72E-05 | 3 |
| 7B95835A-DavidHPeters | 99 | 8 | 3.43E-03 | 3 | 4.48E-02 | 2 | 5.50E-05 | 10 |
| 5F59DCDC-LantPritchett | 57 | 42 | 2.46E-03 | 4 | 4.31E-02 | 14 | 1.91E-05 | 653 |
| 781D4EE0-ZulfiqarABhutta | 93 | 9 | 2.27E-03 | 5 | 4.40E-02 | 4 | 3.15E-05 | 108 |
| 8068F04B-DavidMckenzie | 103 | 7 | 2.27E-03 | 6 | 4.29E-02 | 23 | 6.43E-05 | 5 |
| 7A320C3A-FrankJChaloupka | 77 | 22 | 2.16E-03 | 7 | 4.15E-02 | 118 | 5.03E-05 | 16 |
| | | | | | | | | |
| | | | | | | | | |
| | | | | | | | | |
| 14ABE527-DavidRBangsberg | 92 | 10 | 2.00E-03 | 8 | 4.33E-02 | 13 | 2.62E-05 | 240 |
| 29EA980D-AgnesSoucat | 80 | 20 | 1.95E-03 | 9 | 4.27E-02 | 27 | 3.34E-05 | 83 |
| 75282DF5-GershonFeder | 42 | 57 | 1.91E-03 | 10 | 3.97E-02 | 579 | 2.38E-05 | 325 |

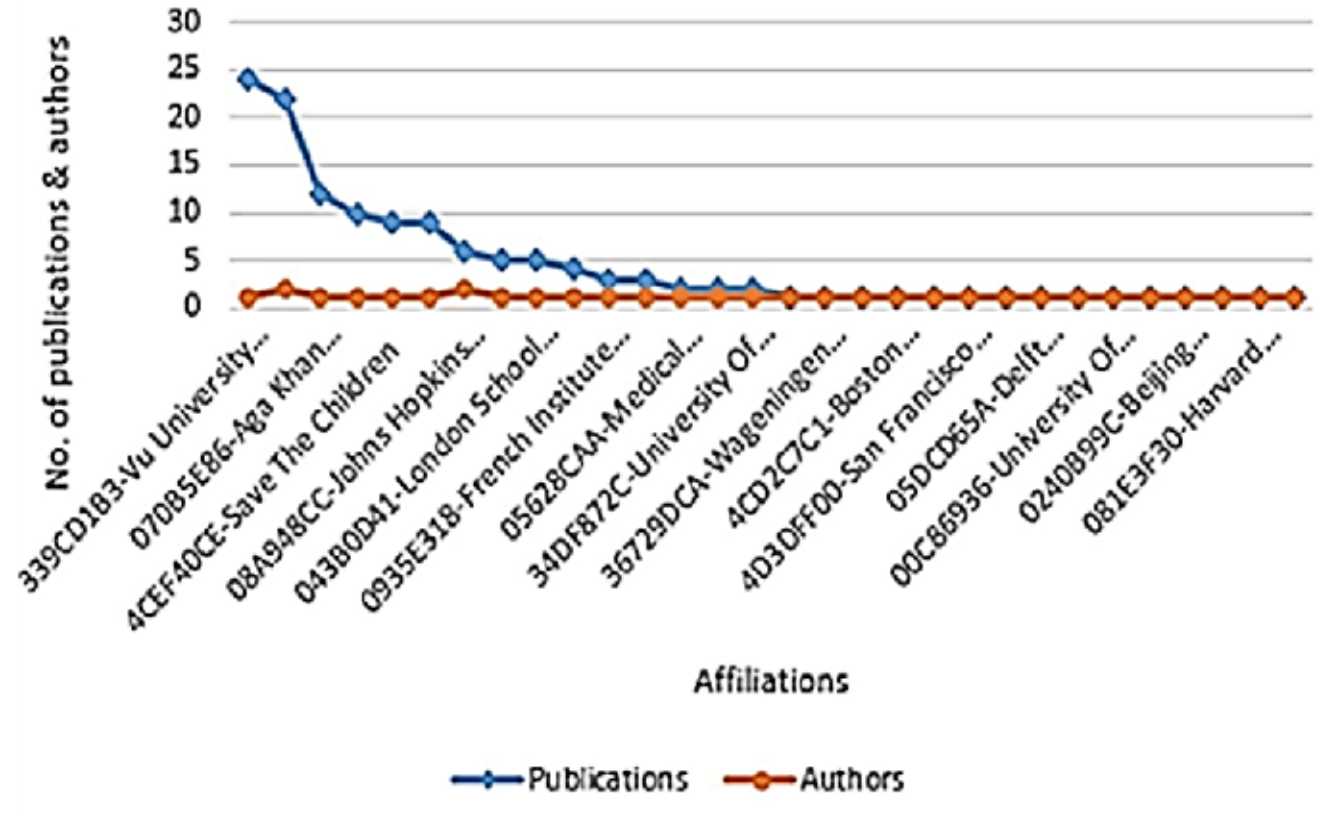

Fig. 12. Institutes and their Publications W.R.T Degree Centralities.

Community of top ten degree researchers and their connected researchers is shown in figure 11. This graph contains 1263 researchers and 1389 collaborations. Average degree of graph is 2.2, network diameter is 6, modularity is 0.829 and there are four connected components. The most productive institute in community of top 10 highest degree researchers of public administrations are the „339CD1B3-Vu University Amsterdam‟, „0A183231-Johns Hopkins Bloomberg School Of Public Health‟, „070B5E86-Aga Khan University‟ and so on as shown in figure 12. „0CAEADF8-Vu‟ has 146 degree and ranked at 3, „7EBE0990RobertEBlack‟ has degree 159 and ranked at 2 and „781D4EE0-ZulfiqarABhutta‟ has 93 degree ranked at 9, are affiliated to „339CD1B3-Vu University Amsterdam‟, „0A183231-Johns Hopkins Bloomberg School Of Public Health‟ and „070B5E86-Aga Khan University‟, respectively.

*B. Ranking Authors based on Betweenness Centrality*

The network diameter of public administrations is 34 and the average path length is 24.833. The highest normalized betweenness is 4.44E-03 and least is zero. The author „7EBE0990-RobertEBlack‟ has collaborated with „0C45A054Diarrhoeal Disease Research Bangladesh‟, „0A183231-Johns School Of Public Health‟, „08A948CC-Johns Hopkins University‟, „070B5E86-Aga Khan University‟, „4CED0A71World Health Organization‟, „05628CAA-Medical Research Council‟, „043B0D41-London School Of Hygiene Tropical Medicine‟, „4CEF40CE-Save The Children‟ and 21 other affiliations, having highest influence and frequent collaboration with other 148 researchers as shown in Table VII. „0CAEADF8-Vu‟ is the second most central researcher and have frequent collaboration as he is ranked 3 in degree centrality measures having 146 degree and affiliated to „339CD1B3-Vu University Amsterdam‟, „3653C029-Vu

Amsterdam‟ „00C86936 -University Of Cantabria‟ and 17 other affiliations. Graph for top ten researchers with respec t to betweenness centrality is shown in figure 13. This graph contains 900 researchers and 944 collaborations.

The network diameter of top 10 betweenness researchers graph is 10, average path length is 4.83, 0.832 is modularity and there are 2 connected components. The most productive institutes in community of top 10 betweenness researchers of public administrations are the „339CD1B3 -Vu University Amsterdam‟, „0A183231-Johns Hopkins Bloomberg School Of Public Health‟, „070B5E86 -Aga Khan University‟, „096500C2-University Of Cape Town‟, and 26 other affiliations. Since they have large number of publications as shown in figure 14. „0CAEADF8 -Vu‟ researcher belongs to „339CD1B3-Vu University Amsterdam‟ and he has 24 publications.

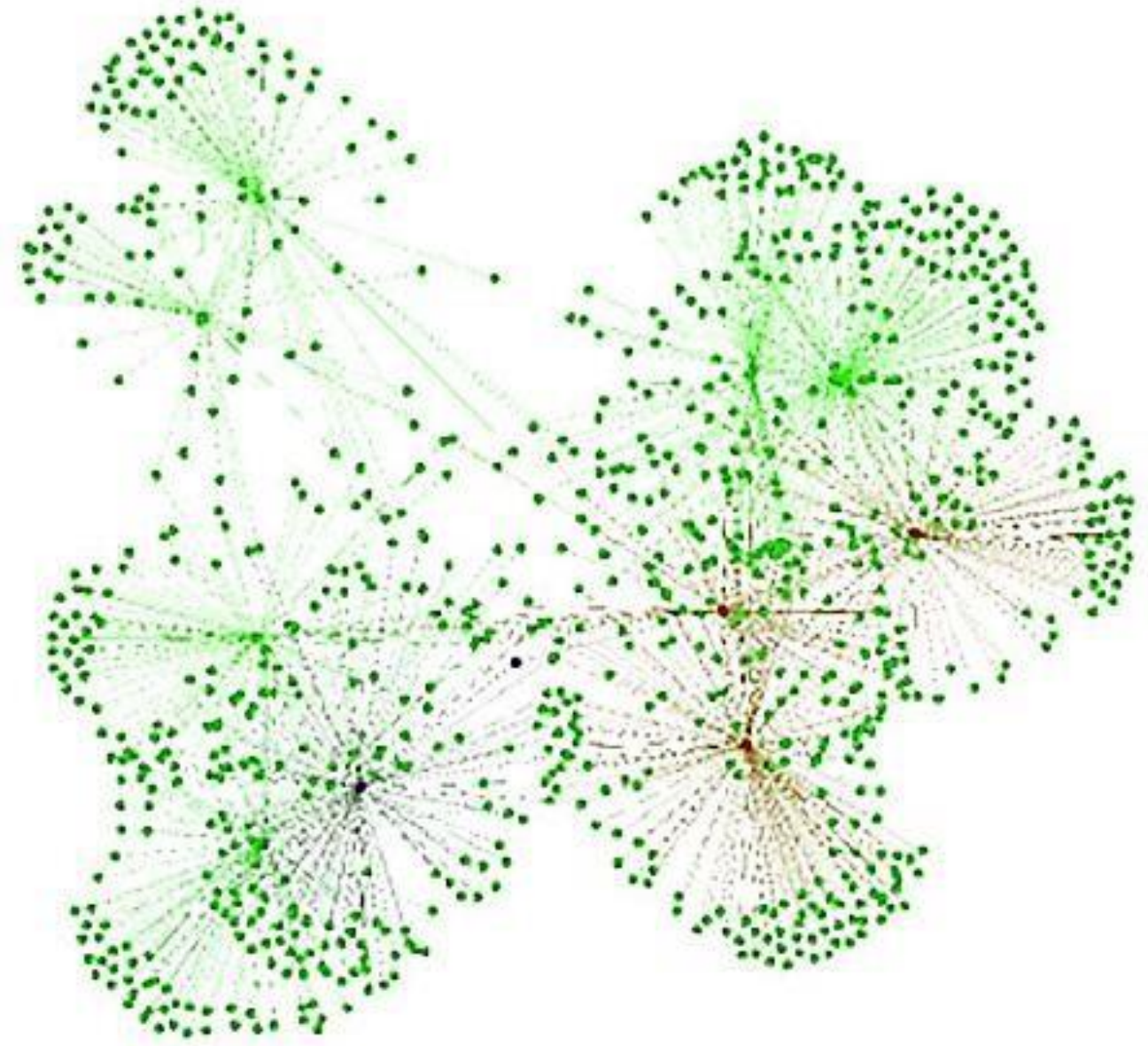

University Medical Center‟, „34DF872C-University Of

Fig. 13. Top 10 Researchers W.R.T Betweenness.

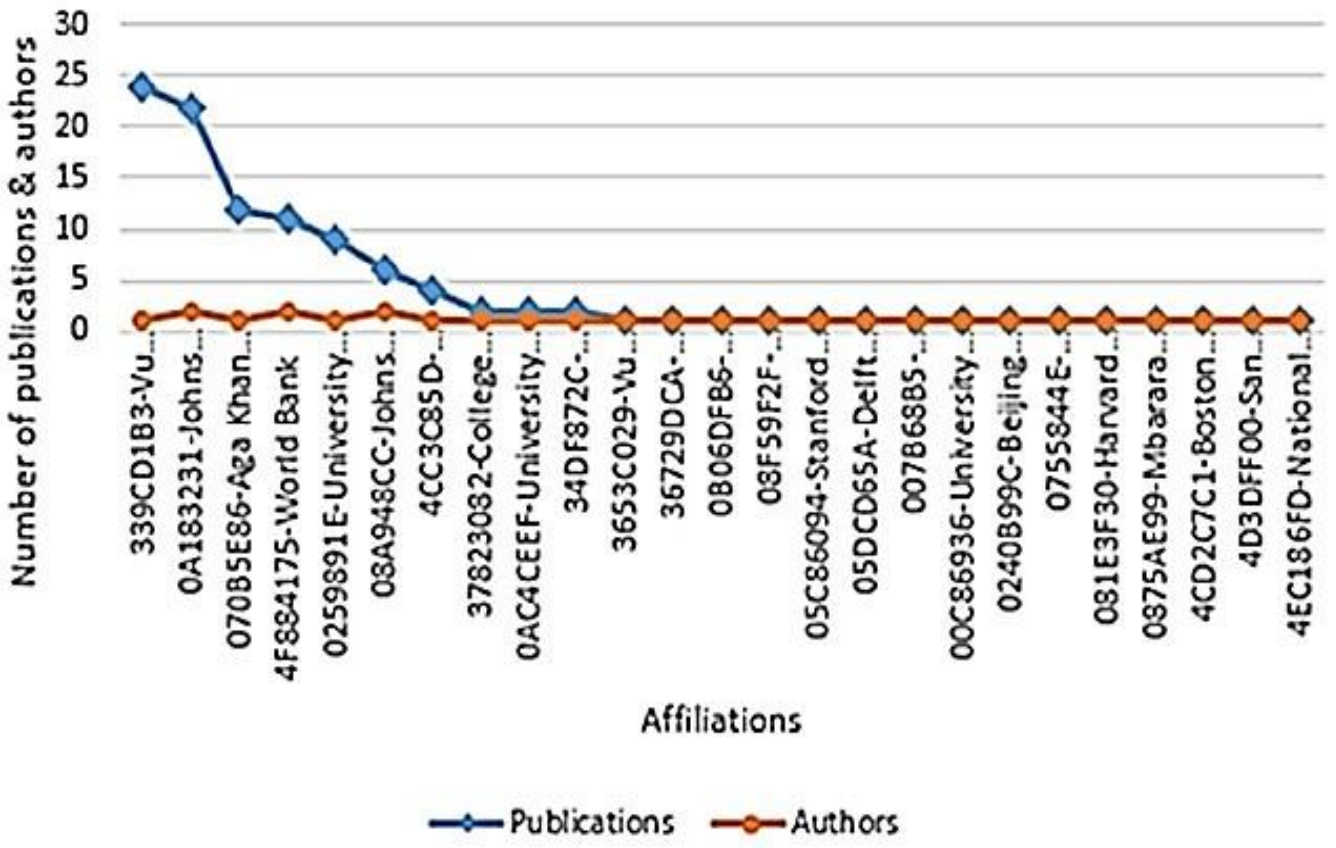

Fig. 14. Institutes and their Publications W.R.T Betweenness Centrality.

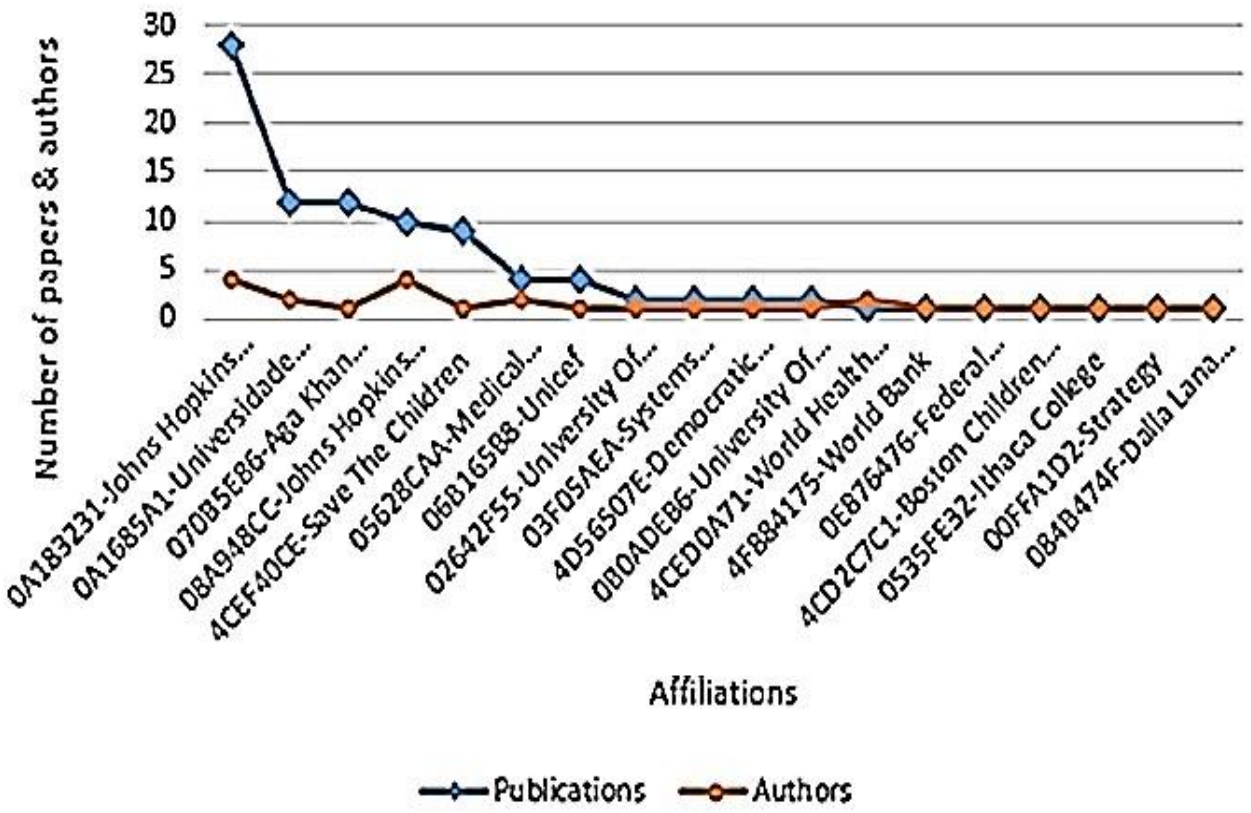

„7B95835A-David H Peters‟ and „7EBE0990-Robert E Black‟ belong to „0A183231-Johns Hopkins Bloomberg School Of Public Health‟ and they have 10 and 20 publications ,respectively.

*C. Ranking Authors based on Closeness Centrality*

The author of public administration field, named „7EBE0990-RobertEBlack‟ is ranked first in closeness, same as in betweenness, as shown in Table VIII. „7B95835ADavidHPeters‟ is the second most rated researcher who is responsible for spreading information frequently to other researchers in a network, since he has ranked 8 in degree centrality measures having 99 degree and prominently affiliated to „08A948CC-Johns Hopkins University‟, „0A183231-Johns School Of Public Health‟, „0992A59E-Makerere University School Of Public Health‟, „0AE9B3CCIndian Institute Of Health Management Research‟ and 12 other affiliations as shown in figure 16.

The most productive institute in community of top 10 closeness researchers of public administrations are the „0A183231-Johns Hopkins Bloomberg School of Public Health Health‟, „0A1685A1-Universidade Federal De Pelotas‟, „070B5E86-Aga Khan University‟, „08A948CCJohns Hopkins University‟, „4CEF40CE-Save The Children‟ and 13 other affiliations as shown in figure 15.

Fig. 15. Institutes with their Publications and Authors of Top 10 Closeness Researchers of Public Administration.

TABLE VIII. TOP 10 AUTHORS RANKING IN PUBLIC ADMINISTRATION ON THE BASIS OF CLOSENESS CENTRALITY

| **Author** | **Degree** | **Rank** | **Betweenness** | **Rank** | **Closeness** | **Rank** | **PageRank** | **Rank** |
|---|---|---|---|---|---|---|---|---|
| 7EBE0990-RobertEBlack | 159 | 2 | 4.44E-03 | 1 | 4.62E-02 | 1 | 5.44E-05 | 11 |
| 7B95835A-DavidHPeters | 99 | 8 | 3.43E-03 | 3 | 4.48E-02 | 2 | 5.50E-05 | 10 |
| 7FD861D8-MickeyChopra | 83 | 17 | 1.12E-03 | 38 | 4.41E-02 | 3 | 3.19E-05 | 105 |
| 781D4EE0-ZulfiqarABhutta | 93 | 9 | 2.27E-03 | 5 | 4.40E-02 | 4 | 3.15E-05 | 108 |
| 130B76BC-VirojTangcharoensathien | 34 | 65 | 7.76E-04 | 99 | 4.39E-02 | 5 | 1.92E-05 | 644 |
| 80FEB1CC-PrabhatJha | 42 | 57 | 1.52E-03 | 15 | 4.38E-02 | 6 | 1.96E-05 | 601 |
| 7DDF7540-RonaldHGray | 74 | 25 | 8.71E-04 | 73 | 4.37E-02 | 7 | 2.67E-05 | 219 |
| 77843A2C-GeoffPGarnett | 44 | 55 | 6.78E-04 | 119 | 4.34E-02 | 8 | 1.28E-05 | 1857 |
| 7D1B2864-NeffWalker | 40 | 59 | 3.21E-04 | 449 | 4.33E-02 | 9 | 1.07E-05 | 2738 |
| 14ABE527-DavidRBangsberg | 92 | 10 | 2.00E-03 | 8 | 4.33E-02 | 10 | 2.62E-05 | 240 |

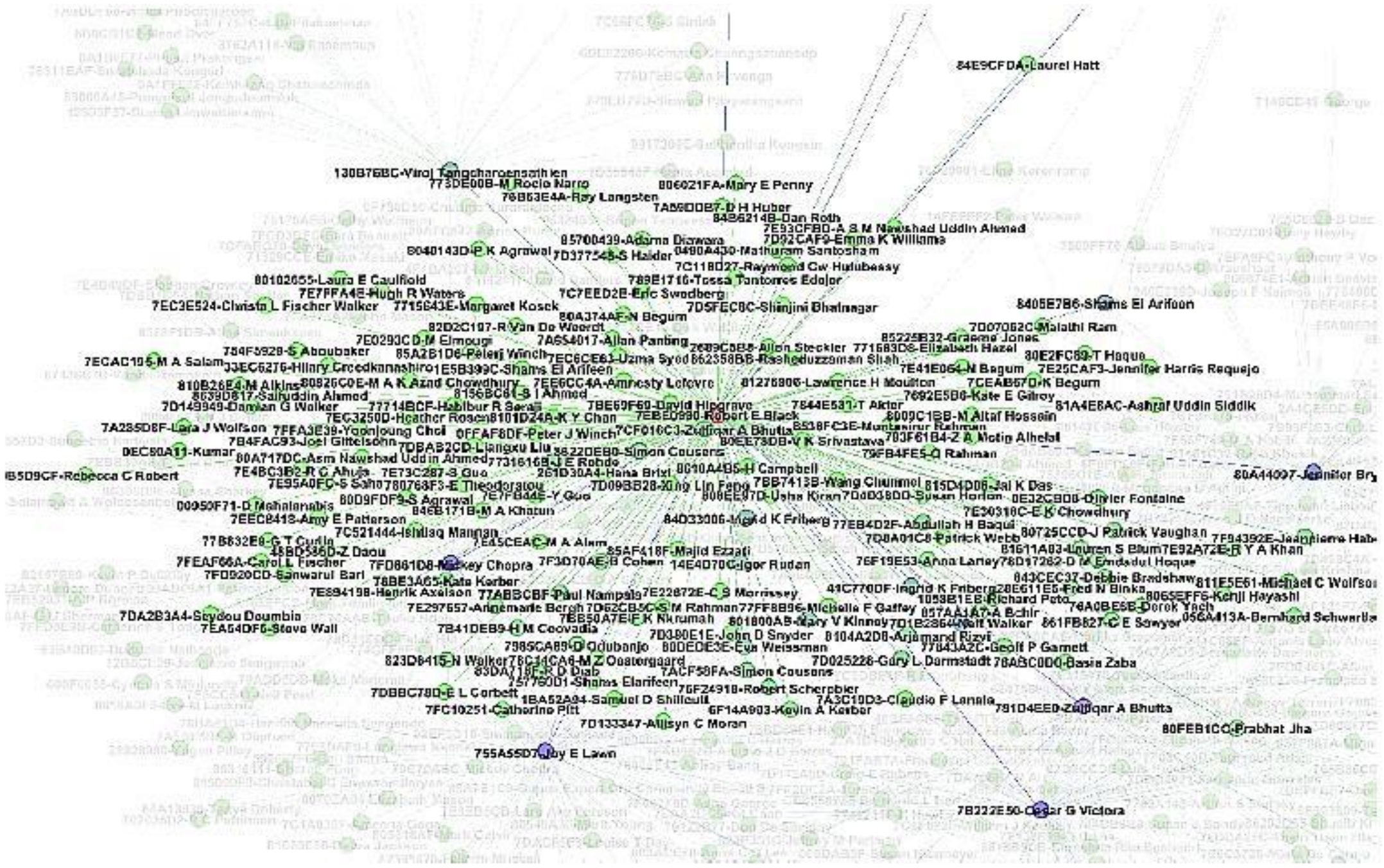

Fig. 16. „7EBE0990-RobertEBlack‟, Author of Public Admininstrations having Highest Closeness Centrality.

TABLE IX. TOP 10 AUTHORS RANKING IN PUBLIC ADMINISTRATION ON THE BASIS OF PAGERANK

| **Authors/Researchers** | **PageRank Value** | **Rank** |
|---|---|---|
| 7F404D7B-PeterDreier | 1.05E-04 | 1 |
| 0CAEADF8-Vu | 8.72E-05 | 2 |
| 7E035912-KristinAMoore | 6.61E-05 | 3 |
| 8068F04B-DavidMckenzie | 6.43E-05 | 4 |
| 7EB811DA-JohnALucas | 6.42E-05 | 5 |
| 618527B9-AntonioEstache | 6.36E-05 | 6 |
| 7D542665-RobertGottlieb | 5.76E-05 | 7 |
| 20CA3DCA-PeterNijkamp | 5.63E-05 | 8 |
| 7B95835A-DavidHPeters | 5.50E-05 | 9 |
| 7EBE0990-RobertEBlack | 5.44E-05 | 10 |

„7EBE0990-Robert E Black‟, „7B95835A-David H Peters‟, „80A44097-Jennifer Bryce‟, and „7DDF7540-Ronald H Gray‟ researchers belong to „0A183231-Johns Hopkins Bloomberg School of Public Health‟ and he have 12, 10, 05 and 03 publications, respectively. „7B222E50-Cesar G Victora‟ and „80A44097-Jennifer Bryce‟ researchers belong to „0A1685A1-Universidade Federal De Pelotas‟ and they have 12 and 1 publications, respectively. „781D4EE0-Zulfiqar A Bhutta‟ researcher belong to „070B5E86-Aga Khan University‟ and this author has 12 publications.

*D. Ranking Authors based on PageRank*

The top ranked researchers who have highest PageRank are shown in Table IX. The author in „Public Administration002F8D8F‟ field named as „7F404D7B-PeterDreier‟ is the researcher who has highest PageRank and has published more than 300 publications by collaborating with 63 researchers related to different fields.

In graph of top 10 PageRank researchers, the most productive affiliation is of „339CD1B3-Vu University Amsterdam‟ with 36 publications as shown in figure 18.

We extracted the graph of top 10 PageRank researchers and their connected researchers as shown in figure 17. This graph contains 645 nodes and 649 edges. Network diameter is 4, average path length is 2.407, modularity is 0.847 and there are seven connected components.

## VII. Discussion

The social network analysis has been widely explored to discover relationship patterns among individuals, teams, groups, societies, communication devices and even among organizations. The study discloses patterns of associations that help in best decision making and better understanding of various patterns in a graph. Analysis study in the domain of co-authorship network helps to identify the dynamic collaboration patterns exist in specific field. We applied centrality measures on two sub fields that is Public Administration and Public Relations of Political Science. We have analyzed just two fields because due to the hardware limitation and the availability of too much nodes where our computer is unable to process more than ten billion nodes. Data is collected from Microsoft Academic Graph. We have taken 102975 papers related to the field of Public Relations and 143831 papers related to Public Administration. For coauthorship network analysis, we selected data that covered time span of 16 years i.e. from 2000 to 2016. We represented the graph in the form of adjacency matrix that is created using Python and R. We considered four common centrality measures for coauthorship network analysis and visualized the centralities and author communities using Gephi and R. Different centrality values for different authors reflect collaborative patterns and trends occurring in 16 years of time span. Analysis on this huge database of public administration and public relation authors discovered the top group of authors who collaborated frequently and diversely in both domains. Some authors hold strong position in a network which shows their strong influence in research collaboration and knowledge sharing.

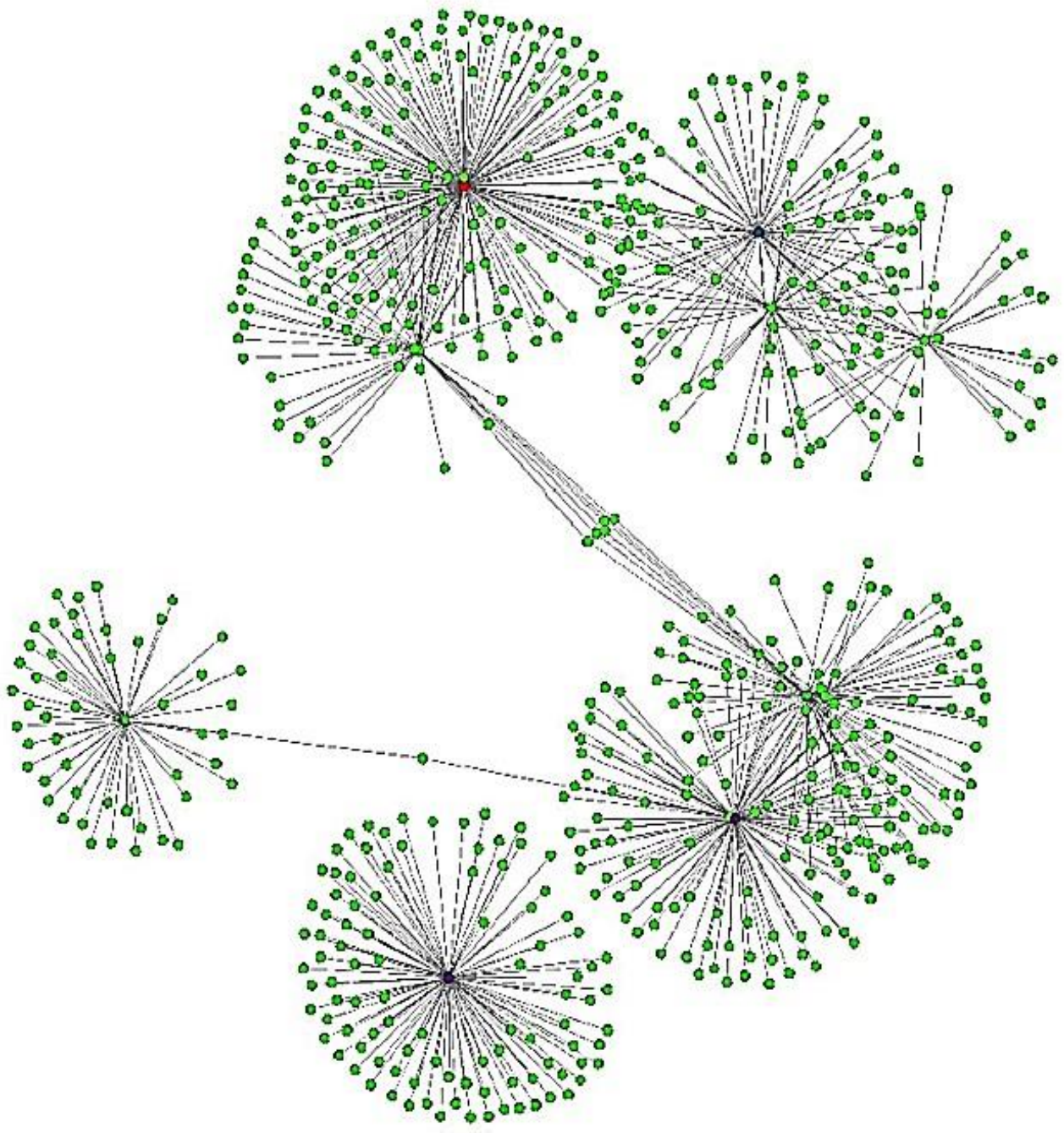

Fig. 17. Graph of Top 10 Authors having Highest PageRank.

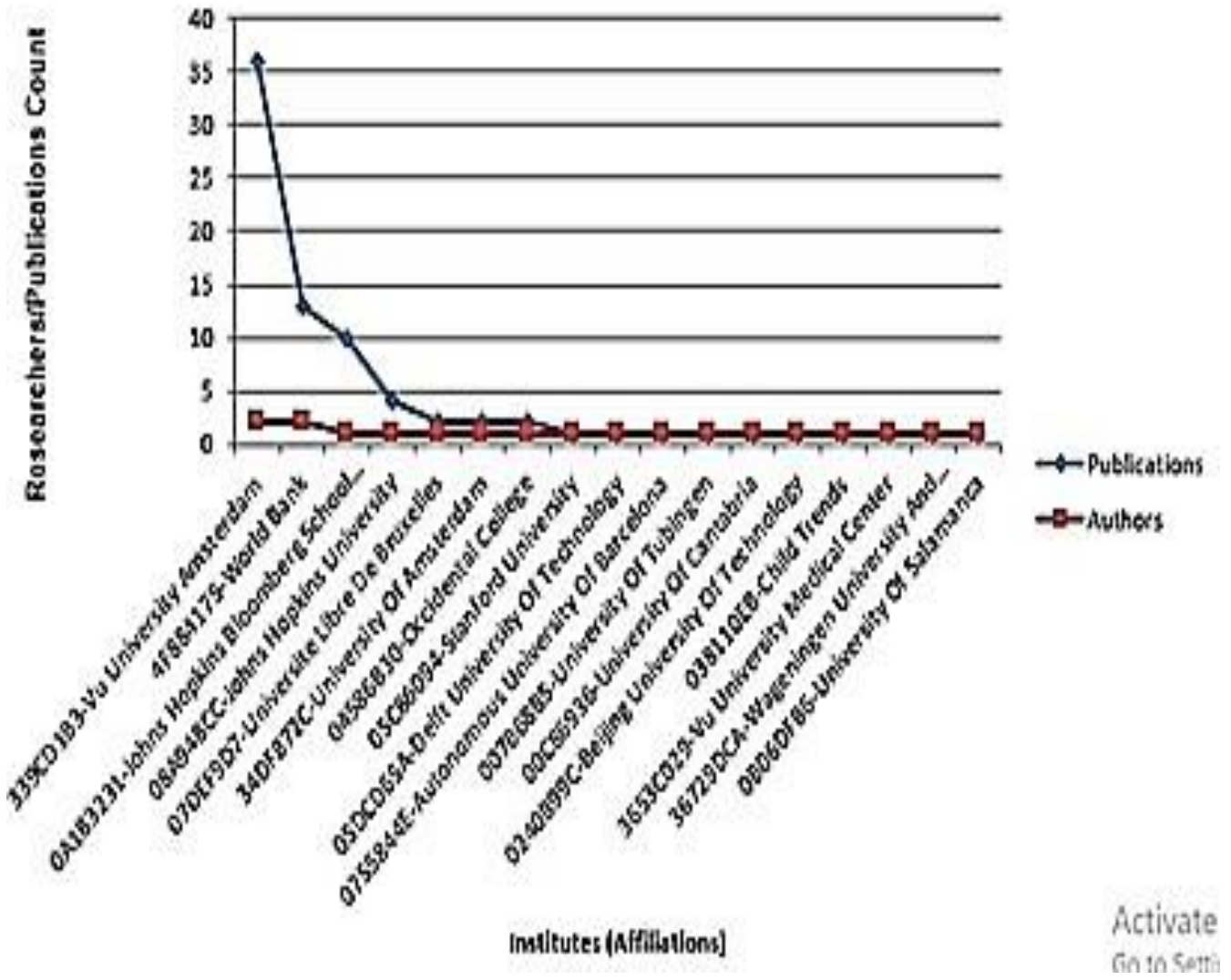

Fig. 18. Institutes and their Publications W.R.T PageRank.

Our analysis is carried out for undirected non-overlapping communities. In future, we will try to carry out an analysis study on directed graph of coauthorship network that will show not only frequent collaboration with co-authors but will also reveal number of publications in relation with other coauthors. There is also a gap to identify the overlapping collaboration among authors because different authors have research contributions in various fields. Other parameters can also be used like impact factor, number of publications and citations count for overlapping community detection to identify and extract the dynamic collaborative patterns in coauthorship network.